# Realizing Blume-Capel Degrees of Freedom with Toroidal Moments in a Ruby Artificial Spin Ice

*Luca Berchialla*[*,1,2], *Gavin M. Macauley*[*,1,2,†], *Flavien Museur*[*,1,2], *Tianyue Wang*[1,2], *Armin Kleibert*[3], *Peter M. Derlet*[1,4] *and Laura J. Heyderman*[*,1,2]

[1] Laboratory for Mesoscopic Systems, Department of Materials, ETH Zurich, 8093 Zurich, Switzerland

[2] PSI Center for Neutron and Muon Sciences, 5232 Villigen PSI, Switzerland

[3] Swiss Light Source, Paul Scherrer Institut, Villigen PSI 5232, Switzerland

[4] PSI Center for Scientific Computing, Theory and Data, 5232 Villigen PSI, Switzerland

ABSTRACT

Realizing exotic Hamiltonians beyond the Ising model is a key pursuit in experimental statistical physics. One such example is the Blume-Capel model, a three-state spin model, whose phase diagram features a tricritical point where second-order and first-order transition lines converge, leading to a coexistence of paramagnetic, ferromagnetic, and disordered phases. Here, we realize an artificial crystal of single-domain nanomagnets, placed on the links of the Ruby lattice, enabling real-space observation of the Blume-Capel degrees of freedom. These Blume-Capel degrees of freedom are represented by the presence, sign and interactions of the toroidal moments that emerge naturally in plaquettes of nanomagnets in the Ruby artificial spin ice. By precisely tuning the lattice parameters of the Ruby artificial spin ice, we demonstrate control over the two-step ordering process of the toroidal moments, whereby there is a high-temperature crossover from a paramagnetic phase to an intermediate paratoroidic regime, followed by a second-order phase transition to a ferrotoroidic ground state. This sequence of toroidal phases and transitions is accurately captured by the Blume-Capel framework and provides a direct realization of a substantial portion of the phase diagram associated with the model. This establishes a new platform for exploring exotic Hamiltonians in terms of artificial spin ice superstructures, here with groups of nanomagnets forming toroidal moments. The success of this mapping paves the way for an entirely new frontier in artificial spin ice: intentionally engineering lattice designs whose effective Hamiltonians mediate unconventional forms of magnetic order, with new behaviors and functionalities.

A central challenge in modern condensed matter physics is the development of experimental platforms that can host novel spin models or even extend existing ones. In many natural materials, the atomistic magnetic order is difficult to tailor and cannot be imaged in real space, leaving the exact details of short and intermediate range magnetic correlations shrouded in uncertainty. At the same time, advances in nanotechnology now make it possible to engineer magnetic interactions on the nanoscale, offering a new frontier for designing functional materials with tailored properties. As such, artificial spin ices[1,2] (ASIs) provide a unique platform to emulate exotic Hamiltonians in a fully accessible, tunable environment, helping to bridge the gap between experiment and theory. By patterning single-domain nanomagnets on different lattices using electron-beam lithography, the interactions in the system can be engineered so that a variety of phase transitions including Ising[3], Potts[4], quadrupolar[4,5] and toroidic[6–8] can be observed.

One spin model that awaits real-space observation is the Blume-Capel model, which features a third spin state, conventionally denoted 0, in addition to the -1 and +1 states common to the Ising model. Specifically, the Hamiltonian of the Blume-Capel model includes a nearest-neighbor spin-spin interaction term, like the Ising model, and an additional term, which controls the population of the 0 state:

$$\mathcal{H}_{\text{B-C}} = -J \sum_{\langle ij \rangle} t_i t_j - \Delta \sum_i t_i^2. \tag{1}$$

The constants $J$ and $\Delta$ set the strength of the nearest-neighbour interaction and the energy difference between spin states $t_i \in \{0, -1, +1\}$, respectively, with $\Delta$ distinguishing the degenerate $\pm 1$ states from the 0 state. This model was independently proposed in 1966 by Blume[9] and Capel[10] to explain critical behavior in two distinct contexts: the first- and second-order magnetic phase transitions in uranium dioxide, and the phase transitions in an Ising system with triplet ions, where the triplet state can be split into a single and a doublet. Since then, the Blume-Capel model has been applied widely to explain a variety of systems including alloys[11], and emergent magnetic monopoles in pyrochlores[12]. Other applications include absorption in two-component gases or liquids[13,14], structural transitions in $VO_2$[15], and the first-order metamagnetic transition in bcc FeRh[16].

While simple to describe, the Blume-Capel model has an extraordinarily rich phase diagram, featuring both first- and second-order phase transitions, as well as crossovers and a tricritical point. To illustrate this, we show the phase diagram of the Blume-Capel model on a triangular lattice as a function of $\Delta/J$ in Figure. 1, as obtained from our Metropolis-Hastings Monte Carlo simulations with the Hamiltonian given by equation (1). Depending on the ratio $\Delta/J$, three different magnetic phases are possible: a high-temperature ternary mixture of all three spin states (+1, -1 and 0), a mixed phase of +1 and -1 spins, or a uniform phase consisting of a single spin type.

Here, we introduce a novel ASI based on the Ruby lattice in order to realize the degrees of freedom associated with the Blume-Capel model. Specifically, we construct a system where we can identify three distinct spin states—analogous to +1, -1, and 0—and two effective interactions, corresponding to $\Delta$ and $J$. We fabricate a lattice with elongated single-domain nanomagnets, whose magnetic state can be described by an Ising degree of freedom. The magnetic moment associated with each nanomagnet—its "macrospin"—can be readily imaged, for example using magnetic force microscopy[8] (MFM) and synchrotron x-ray photoemission electron microscopy[17] (PEEM).

The specific arrangement of nanomagnets defines the interactions within the system, a fact that has been exploited to observe phenomena such as emergent magnetic monopoles[18,19], collective dynamics[20,21], vertex frustration[22–24] and phase transitions[4,17,25] in ASI. Here, we place the single-domain nanomagnets on the links of the Ruby lattice, which is a two-dimensional network of hexagons and triangles, as shown in Figure 2a. Due to the dipolar interaction, it is energetically favorable for the macrospins associated with neighboring nanomagnets to align head-to-tail. The low-energy states of the Ruby ASI thus feature flux-closed hexagons and triangles, where all the nanomagnets lining a given shape form loops of head-to-tail macrospins. These flux-closed superstructures—in effect, emergent toroidal moments—act as the +1 and -1 spin state associated with our effective Blume-Capel variable. The third spin state, the 0 state, corresponds to those hexagons or triangles where the associated macrospins do not all align head-to-tail or, equivalently, where there is no fully-formed toroidal moment.

By tuning the two lattice parameters that define the Ruby ASI (Figure 2a-d), we demonstrate precise control over the toroidal ordering in the system. In this way, we can reach the ferrotoroidic ground state, defined by a uniform alignment of toroidal moments, either (i) through a single phase transition (Figure 3a, labelled $I_{Tr} \approx I_{Hex}$) or (ii) through a crossover followed by a phase transition (Figure 3a, labelled $I_{Tr} > I_{Hex}$ or $I_{Tr} < I_{Hex}$) on decreasing the temperature. Interestingly, in the second scenario, the system is in a paratoroidic regime between the crossover and phase transition. In this regime, individual hexagons or triangles have formed toroidal moments, but there is no long-range correlation between them. We show that the Blume-Capel degrees of freedom can be used to accurately describe these toroidic phase transitions and crossovers. With micromagnetic estimates for the effective $J$ and $\Delta$, along with extensive MFM and PEEM measurements of both frozen and thermally active Ruby ASIs, we explore an unconventional part of the Blume-Capel phase diagram, as shown by the region shaded in red in Figure. 1. While the Ruby ASI cannot exhibit negative values of $\Delta$, which are associated with the tricritical point and the first order phase transition, we instead explore this previously uncharted region, offering new insights into the critical behavior of the phase diagram.

**Controlling the ordering in the Ruby ASI through Two Lattice Parameters**

The Ruby—or Rhombitrihexagonal—lattice is one of the 11 Archimedean tilings[26] where the same polygons surround each vertex. In the Ruby lattice, when travelling around a vertex, one comes across a triangle, a rectangle, a hexagon, then a final rectangle. The Ruby lattice is therefore also referred to as the 3,4,6,4-tiling, where each number indicates the number of sides of successive polygons at a vertex[27].

We define the Ruby lattice through two independent lattice parameters, $a$ and $b$, which represent the side lengths of the triangles and hexagons in the lattice, respectively. The unit cell of the Ruby ASI is highlighted in yellow in Figure 2a, and features 12 nanomagnets. One can consider the Ruby ASI as being composed of a network of interpenetrating triangles (of side length $a$) and hexagons (of side length $b$), or equivalently, composed of a lattice of rectangles of (of dimensions $a \times b$). We will describe the behavior of the lattice focusing only on the triangles and the hexagons, but note that considering just the rectangles provides an equivalent way of grouping the nanomagnets of the Ruby ASI.

Using electron beam lithography, we prepare several Ruby ASIs, each with a different lattice parameter ratio ($a/b$), by separately increasing the value of each lattice parameter starting from $a_{min}$ = 695nm and $b_{min}$ = 535 nm in 20 nm and 15 nm steps, respectively. Since each lattice parameter controls the side length of triangles or hexagons, each of them also determines the interaction strength between nanomagnets lying in a triangle ($I_{Tr}$) or a hexagon ($I_{Hex}$). Increasing each lattice parameter independently from ($a_{min}$,$b_{min}$), at which ($I_{Tr} \approx I_{Hex}$), produces two series of arrays, one where $I_{Tr}$ becomes increasingly smaller than $I_{Hex}$ and vice versa.

Depending on the lattice parameter ratio (*a*/*b*), which directly sets the ratio of interaction strengths ($I_{Tr}/I_{Hex}$), we identify three different regimes, illustrated by the lattices in Figure 2b-d, which are determined by the shape (triangle or hexagon) whose macrospins first align head-to-tail to form flux-closed loops upon cooling:

When $I_{Tr} > I_{Hex}$ (Figure 2b), the macrospins within a triangular plaquette align head-to-tail first because the nanomagnets in a triangle are tightly packed and therefore strongly coupled. However, the distance between the centres of triangular plaquettes is large or, equivalently, the hexagons have a large side length.

When the two interaction strengths are approximately equal, $I_{Tr} \approx I_{Hex}$ (Figure 2c), both the triangles and the hexagons are closely packed, and flux closure of both shapes occurs more-or-less simultaneously, depending on the local environment.

When $I_{Tr} < I_{Hex}$ (Figure 2d), the macrospins within a hexagonal plaquette align head-to-tail first because the nanomagnets in a hexagon are tightly packed and therefore strongly coupled. However, the distance between the centres of hexagonal plaquettes is large or, equivalently, the triangles have a large side length.

When the macrospins in a triangle or a hexagon form a flux-closed loop, we can associate a fully-formed toroidal moment to that shape. The sign of the toroidal moment is determined by the sense of circulation. In figures, we represent these fully-formed toroidal moments by coloring the associated plaquettes green (positive toroidal moment; clockwise circulation) or pink (negative toroidal moment, anticlockwise circulation) as illustrated in Figure 2e.

**Phase diagram with two-step ordering**

On varying the lattice parameters, the Ruby ASI can host four different magneto-toroidal phases as a function of temperature, which are illustrated in Figure 2e. On cooling, the Ruby ASI can relax to the ferrotoroidic ground state (i) from the high temperature paramagnetic phase (iv), either directly or by first traversing an intermediate paratoroidic regime (ii, iii). In ferrotoroidic ground state (i), all triangular and hexagonal plaquettes host toroidal moments of the same sign. In the paratoroidic regimes (ii, iii), either all triangular plaquettes (ii) or all hexagonal plaquettes (iii) have toroidal moments, but there is no long-range correlation between toroidal moments. We have calculated the phase diagram using Monte Carlo simulations (Figure 3a), which allow us to describe how the Ruby lattice can access the different magneto-toroidal phases.

To construct the phase diagram of the Ruby ASI, we determine the specific heat capacity $c_V$ from Monte Carlo simulations with a point-dipolar Hamiltonian using both single spin flips and loop moves for hexagons and triangles. Further details are given in the Methods section. A waterfall plot of $c_V$ as a function of temperature is shown in Figure 3a, going from $I_{Tr} > I_{Hex}$ (top, yellow curves) through to $I_{Tr} \approx I_{Hex}$ (middle, green curves), and finally to $I_{Tr} < I_{Hex}$ (bottom, blue curves).

While the apparent nature of the phase transition is the same for all lattice parameters, and appears compatible with a second-order 2D Ising transition, as suggested by our finite size scaling analysis (see Methods), there is a striking difference in short-range physics as the two lattice parameters are varied. When $I_{Tr} \approx I_{Hex}$, there is a single peak in the heat capacity, which corresponds to the phase transition at which the toroidal moments associated with both triangles and hexagons are formed and the ferrotoroidic order is established in a single step. The ferrotoroidic order can be observed in the thermally annealed magnetic states when $I_{Tr} \approx I_{Hex}$, as shown in the PEEM images in the middle panel of Figure 3b. Accordingly, the magnetic structure factor for the ferrotoroidic ground state displays sharp magnetic Bragg peaks (middle panel of Figure 3c).

However, when either $I_{Tr} > I_{Hex}$ or $I_{Tr} < I_{Hex}$, two features appear in $c_V$. At high-temperatures, there is a broad peak that occurs at a temperature on the order of the nearest-neighbor interaction strength $k_B T/D \approx 1$ (Figure 3a), where $k_B$ is the Boltzmann constant and $D$ is the dipolar constant. The quantity $D$ depends on both the saturation magnetization and volume of the nanomagnets, as well as the scale of their separation. The broad peak at $k_B T/D \approx 1$ represents a crossover to a paratoroidic regime associated with the flux closure of the nanomagnets in the tightly-packed shapes. We interpret this broad peak as a crossover because there is no global symmetry breaking of either the spin ensemble or the toroidal moment ensemble. This can be seen because there is no correlation between the sign of adjacent toroidal moments. However, there is a local symmetry breaking on the level of individual triangular or hexagonal loops, which choose either clockwise or anticlockwise circulation. This behavior is consistent with the thermally annealed magnetic states shown in the PEEM images in the left and right panels of Figure 3b, where we observe that toroidal moments have formed only in the tightly-packed plaquettes.

When the temperature is lowered further, a second sharp peak is observed in $c_V$ (Figure 3a, for $I_{Tr} > I_{Hex}$ or $I_{Tr} < I_{Hex}$). As the system is cooled through this peak, the mirror symmetry of the spin ensemble and, hence, the toroidal moment ensemble is broken, leading to an alignment of the toroidal moments of the tightly packed shapes. At the same time, the toroidal moments associated with the other shape form and align to the ones already present in the lattice. This brings the lattice into the same ferrotoroidic ground state as that for $I_{Tr} \approx I_{Hex}$ (PEEM image in middle panel of Figure 3b). The position of this peak can be moved to lower temperatures by increasing the relevant lattice parameter.

The corresponding magnetic structure factors in the paratoroidic regime exhibit diffuse Bragg peaks. These are arranged on a hexagonal lattice, reflecting the fact the unit cell of the ruby lattice (yellow nanomagnets in Figure 2a) is arranged on a triangular lattice in real space. The diffuse nature of the scattering is expected because only a subset of spins have ordered, corresponding to one set of fully formed toroidal moments, and these toroidal moments are not correlated with each other. The structure factor for $I_{Tr} > I_{Hex}$ is more diffuse than the one for $I_{Tr} < I_{Hex}$. In the latter case, the high-intensity regions (dark contrast in Figure 3c) are distinctly separated, whereas in the former, they merge into broader, less well-defined features. This difference reflects the fact that the critical temperature for $I_{Tr} < I_{Hex}$ is higher. At the same effective temperature, the $I_{Tr} < I_{Hex}$ system is closer to its critical transition and, therefore, more ordered, producing sharper features in the magnetic structure factor.

At this point, we emphasize that the Ruby ASI has the same ferrotoroidic ground state irrespective of the chosen lattice parameter. However, by changing the lattice parameter ratio $a/b$, and thus the ratio of interaction strengths, one can control whether the paratoroidic regimes exist and their nature, i.e. whether they are formed by toroidal moments of the hexagons or the triangles.

**Experimental signatures of the two-step ordering process and toroidic phases**

To obtain experimental signatures of the phase diagram in Figure 3a, we determine the magnetic state of as-grown configurations using MFM. In addition, we calculate $t_{H+}$, $t_{T+}$ ($t_{H-}$, $t_{T-}$) as the fractional populations of positive (negative) fully-formed toroidal moments associated with a hexagon and a triangle plaquette, respectively, and these are shown for different lattice parameter ratios in Figure 4b-c. We also define an intensive (or system size independent) ferrotoroidic order parameter $\Phi$ as shown in Figure 4d, which is a measure of the extent to which long-range ordering of toroidal moments is established through

$$\Phi = \frac{1}{2}(|t_{H+} - t_{H-}| + |t_{T+} - t_{T-}|). \tag{2}$$

As-grown and annealed configurations are comparable since an ASI undergoes an effective thermal annealing process during the initial stages of magnetic material deposition[28,29]. For this reason, all as-grown configurations have roughly the same effective temperature (Figure 4a), because the single-nanomagnet blocking temperature $T_B$, which is the highest temperature at which the macrospin is unlikely to reverse in a given timescale, is determined to first order by the size of the nanomagnets[30]. By imaging many lattices of varying $(a,b)$, we traverse the phase diagram along a line of approximately constant effective temperature (dashed blue line in Figure 1 and vertical dashed black line in Figure 3a). Both paratoroidic and ferrotoroidic regimes can be accessed because we fabricate several Ruby ASIs with increasing nearest-neighbor interactions within triangles ($I_{Tr}$) and hexagons ($I_{Hex}$), with the nearest-neighbor interaction going from below to above the single-nanomagnet blocking temperature.

For $(a,b)$ close to $(a_{min}, b_{min})$, indicated in Figure 4 by the vertical dashed line, both $I_{Tr}$ and $I_{Hex}$ are large enough such that the critical temperature of the resulting phase transition to the ferrotoroidic ground state exceeds $T_B$. This implies that both hexagonal and triangular plaquettes host fully-formed toroidal moments in the as-grown configurations as indicated by $t_{H\pm} \sim 1$ and $t_{T\pm} \sim 1$ in Figure 4b,c. Moreover, there is also a ferrotoroidic order as suggested by $\Phi \sim 1$ in Figure 4d.

As soon as either lattice constant is increased, we first observe a reduction of the ferrotoroidic order ($\Phi < 1$, Figure 4d). Then, as either $I_{Tr}$ or $I_{Hex}$ is reduced further still, the phase transition temperature to the ferrotoroidic ground state also reduces, eventually becoming lower than $T_B$ and the corresponding population of fully-formed toroidal moments ($t_{H\pm}$ or $t_{T\pm}$ in Figure 4b,c) decreases. When only one type of fully-formed toroidal moment is present, the system reaches a paratoroidic regime of uncorrelated toroidal moments ($\Phi \sim 0$, Figure 4d).

**Blume-Capel degrees of freedom for toroidal moments and phase transitions**

We have established the existence of a ferrotoroidic ground state in the Ruby ASI (Figure 2e, i), which can be accessed either directly from the paramagnetic phase (Figure 2e, iv) or through an intermediate paratoroidic regime (Figure 2e, ii and iii) with toroidal moments on the closely-packed plaquettes. The route to the ground state depends on the ratio of the lattice parameters. For the nanomagnets lining a given plaquette, there are three scenarios: the nanomagnets can all align head-to-tail in a clockwise sense, resulting in a positive toroidal moment; they can all align head-to-tail in an anticlockwise sense, leading to a negative toroidal moment; or they may not all align head-to-tail, leaving the plaquette without a fully-formed toroidal moment.

We now show that, interestingly, these three possibilities for plaquettes with closely-packed nanomagnets can be mapped onto the three states of the spin variable $t$ in the Blume-Capel model. In particular, using Blume-Capel degrees of freedom, we can describe the behaviour of the hexagonal toroidal moments when $I_{Tr} < I_{Hex}$ ($a/b > a_{min}/b_{min}$) and of the triangular toroidal moments when $I_{Tr} > I_{Hex}$ ($a/b < a_{min}/b_{min}$). As a reminder, the Hamiltonian of the Blume-Capel model, given in equation (1) and Figure 5a, is composed of two terms: a nearest-neighbour interaction of strength $J$, and an anisotropy term, which is the energy difference $\Delta$ between the 0 and the $\pm 1$ states. The spin variable $t$ in the Blume-Capel Hamiltonian represents the toroidal moments of the closely-packed plaquettes.

In the Ruby ASI, we can define an effective nearest-neighbor interaction $J$ between adjacent toroidal plaquettes by comparing the energies of two configurations, as shown in Figure 5a for hexagonal toroidal moments. One configuration is a patch of ground state in which both hexagons (or both triangles) have toroidal moments of the same sign. In the second configuration, the sign of one of the toroidal moments is reversed by reversing the directions of its associated macrospins, while all of the other macrospins are left the same. The energy difference between these two configurations is $2J$. In a similar way, $\Delta$ is defined as the difference between the average energy of excited states ($E_{\text{High}}$) and the ground state ($E_{\text{Low}}$) of a hexagonal (or triangular) plaquette as shown in Figure 5a.

A suitable order parameter for the Blume-Capel model is

$$\Psi = (n_{+1} + n_{-1}) + |n_{+1} - n_{-1}|, \qquad (3)$$

where $n_{+1}$ and $n_{-1}$ are the fractional populations of toroidal moments in the $t = +1$ and $t = -1$ states, respectively. The order parameter $\Psi = 0$ when all toroidal moments are in the $t = 0$ state; $\Psi = 1$ when the toroidal moments are equally distributed between $t = +1$ and $t = -1$ (paratoroidic regime); and $\Psi = 2$ when all the toroidal moments have all the same state of $t = +1$ or $t = -1$ (ferrotoroidic state).

In Figure 5b, we display $\Psi$ over the temperature range from $10^{-2}$ to $10^{2}$ as a function of $a/b$ for Monte Carlo simulations of the Ruby ASI with one Ising variable per macrospin (upper panel), and Monte Carlo simulations using Blume-Capel degrees of freedom (lower panel) with $J(a,b)$ and $\Delta(a,b)$ values. We focus on the portion of $(a,b)$-space where $b$ is kept constant at $b_{\text{min}}$ while $a$ is increased, i.e., the nanomagnets within each hexagon are closely packed, but the separation between the hexagons increases as $a$ is increased. This corresponds to the interaction strength regime $I_{\text{Tr}} < I_{\text{Hex}}$. In the phase diagram for the Ruby ASI, given in the upper panel of Figure 5b, we observe that, at high temperatures, $\Psi = 0$ (blue region) since only a few hexagonal toroidal moments have formed. As $a/b$ increases, so does the temperature range over which the paratoroidic order is present where $\Psi = 1$ (green region). At low temperatures, the ferrotoroidic phase dominates and $\Psi = 2$ (yellow region).

Monte Carlo simulations using Blume-Capel degrees of freedom (lower panel in Figure 5b) yield similar results, with comparable temperatures of the phase transitions and crossovers. The value that the order parameter takes in the different regions is similar, except at high temperatures where the Blume-Capel model has a mixture of 0, +1 and -1 states, each of which appear with an equal probability of 1/3, yielding $\Psi = 2/3$. In contrast, the probabilities of those states in the Ruby ASI are 62/64, 1/64 and 1/64 for toroidal moments on hexagonal plaquettes and 6/8, 1/8 and 1/8 for toroidal moments on triangular plaquettes. For this reason, we also simulate a modified Blume-Capel model (Supplementary Figure 7c and Figure 8c), where we adjust the probabilities of the three states from 1/3 to match those of the toroidal moments on the Ruby ASI in order to better capture its high-temperature behavior. The Blume-Capel model with modified probabilities provides the best agreement with Monte Carlo simulations of the dipolar Ruby ASI. Further details regarding the Monte Carlo simulation of the Blume-Capel model are contained in the Methods.

Viewed in terms of the Blume-Capel degrees of freedom and for large $\Delta/J$, the behavior of the Ruby ASI upon cooling is dictated at first by $\Delta$, which leads to the formation toroidal moments, and then by $J$, which promotes the toroidal moments to align in the same direction. Since the phase transition from the paratoroidic regime to the ferrotoroidic ground state is analogous to the ferromagnetic transition of Ising spins on a triangular (or hexagonal) lattice, the critical temperatures as a function of $J$ for such transitions can determined exactly. For $a/b > a_{min}/b_{min}$, the hexagon toroidal moments are arranged on a triangular lattice, for which the Ising critical temperature is $T_{Crit} = 4J/\log(3)$[31]. Instead, for $a/b < a_{min}/b_{min}$, the triangle toroidal moments are on a hexagonal lattice, resulting in $T_{Crit} = 2J/\log(2 + \sqrt{3})$ [31]. This provides an even simpler framework, the Ising model, to describe the paratoroidic to ferrotoroidic transition for large $\Delta/J$.

**Conclusions**

We have characterized the magnetic phases of the Ruby ASI, where two independent lattice parameters can be adjusted to tune the nearest-neighbor interactions between nanomagnets in the triangular and hexagonal plaquettes. By adjusting the side lengths of the triangles or hexagons, we can control whether the system reaches its ground state through a single phase transition, or through a crossover followed by a phase transition. In the paratoroidic regime that occurs after the crossover, the macrospins of nanomagnets in the most closely-packed shapes have aligned to form head-to-tail loops. As evidence of these different ordering pathways in the Ruby ASI, we have experimentally tracked the fraction of toroidal moments across a wide range of lattice parameters.

Interestingly, we find that it is possible to map the presence and sign of toroidal moments in the Ruby ASI to a three-state spin variable. In doing so, we simplify the dipolar interactions between individual macrospins in the ASI, showing that they can be well represented by a short-range effective model based on emergent toroidal moments. This provides for the first time real-space experimental observations of the Blume-Capel degrees of freedom, based on hexagon and triangle toroidal moments on 2D triangular and hexagonal lattices, respectively. This provides a new route to the direct observation of thermally-activated dynamics and non-equilibrium behaviour in areas of the Blume-Capel phase diagram, which has so far proven to be extremely challenging in bulk systems. While we cannot access the tricritical point in the Ruby ASI, we envisage that the tricritical point predicted by the Blume-Capel model could be explored by other ASI geometries designed to have superstructures, or toroidal moments, with antiferromagnetic interactions.

This work lays the foundation for exploring exotic spin Hamiltonians through the precise engineering of ASI lattices. In this new paradigm, the spin variables of these Hamiltonians are reinterpreted in terms of the collective magnetic state of a group of nanomagnets, in effect a superstructure. This opens new possibilities for reconfigurable magnetic metamaterials with programmable interactions, thus providing tunable critical behavior and excitations. Such control would allow artificial spin systems to mimic models with multi-state variables[32,33], or directional constraints[34], offering potential pathways toward spin-based logic[35–38] or neuromorphic architectures[39,40]. It may even be possible to engineer an ASI with effective directional bonds between the superstructures, which is evocative of the highly coveted Kitaev model[41].

## Methods

### Sample fabrication

Arrays of Ruby ASIs are fabricated on a 10 × 10 mm$^2$ silicon substrate. The base nanomagnet in each of our patterns is a stadium-shaped permalloy island with lateral dimensions of 450 nm by 150 nm. The minimum lattice parameters ($a_{min}$, $b_{min}$) are selected to ensure that the minimum edge-to-edge distance between adjacent nanomagnets in both triangles and hexagons is the same, namely ∼15 nm. The first step in the fabrication process is to spincoat the substrate with a PMMA (polymethyl methacrylate) resist. The resist is then patterned using a Vistec EBPG 5000PlusES electron beam writer at 100 keV accelerating voltage and developed in a 1:3 mixture of methyl isobutyl ketone and isopropanol, before being rinsed with isopropanol, and spin-dried. The permalloy ($Ni_{80}Fe_{20}$) was deposited via thermal evaporation under a base pressure of 1x10$^{-6}$ mbar and capped, without breaking vacuum, with a 2 nm aluminium layer to prevent oxidation. An ultrasound-assisted lift-off process in acetone removed the undeveloped resist, leaving only the patterned nanomagnets on the sample. For the samples imaged with PEEM, a permalloy wedge thickness going from 0 nm to 7 nm over a distance of 8 mm was deposited by means of a motorised sliding shutter. For the Ruby ASI samples imaged with MFM, the permalloy thickness was 20 nm.

### Synchrotron x-ray photoemission electron microscopy

X-ray photoemission electron microscopy (PEEM) measurements were performed at the Surfaces / Interfaces: Microscopy (SIM) beamline of the Swiss Light Source, Paul Scherrer Institute, Villigen, Switzerland. The magnetic contrast is obtained through the x-ray magnetic circular dichroism (XMCD) effect by shining left and right circularly polarised x-rays at the iron L$_3$ edge onto the sample at a 16° grazing incidence angle. The XMCD contrast is proportional to the scalar product of the magnetization with the x-ray propagation vector. This results in a bright (dark) contrast for a nanomagnet with the magnetisation pointing towards (away from) the x-ray propagation vector. Since the Ruby ASI has 6 different nanomagnet directions separated by 30°, the optimal angle between the x-rays and any nanomagnet long axis is 15°. For our single-domain nanomagnets, where the magnetization points in one of two directions parallel to the nanomagnet long axis, this ensures that the magnetization is never perpendicular to the x-ray propagation vector, which would result in zero XMCD contrast.

### Monte Carlo simulations

Monte Carlo simulations were performed using the Metropolis-Hastings algorithm. For the Ruby ASI, simulations were carried out on a lattice of 10 × 10 unit cells, giving a total of 1200 spins, with no boundary conditions. The observables were averaged across 100 individual simulations. For a given ($a,b$), we determine the six distinct interactions within a unit cell using micromagnetic simulations, as shown in Supplementary Figure 2. All other interactions are calculated with the point dipole approximation and scaled appropriately to match with the micromagnetic simulations. We implemented loop moves for flux-closed hexagons and triangles in order to speed-up the simulations. For the three extreme cases of $I_{Tr} > I_{Hex}$, $I_{Tr} \approx I_{Hex}$ and $I_{Tr} < I_{Hex}$, we also perform a finite size scaling analysis, which reveals a reasonable collapse of the heat capacities when assuming the relevant critical exponent for a second-order 2D Ising transition ($\nu = 1$), suggesting that it belongs to that universality class. The finite size scaling graphs are given in Supplementary Figure 3. The critical temperature of the phase transition is determined with the Binder fourth-order cumulant analysis using an order parameter calculated as the square root of the intensity of the magnetic structure factor in the two diffraction peaks appearing at $Q_{xy} = \{ [4(a+b\sqrt{3}), 0]; [0, 2(\sqrt{3}a + 3b)] \}$.

Monte Carlo simulations of the Blume-Capel model have been performed on a triangular lattice of 2500 spins and also on a hexagonal lattice of 2450 spins with no boundary conditions. The simulations on the triangular (hexagonal) lattice correspond to the toroidal moments associated with hexagonal (triangular) plaquettes arranged on a triangular (hexagonal) lattice in the Ruby ASI.

In the Blume-Capel model, each state (+1, -1 and 0) of the variable $t_i$ is equally likely with a probability of 1/3.

In the modified Blume-Capel model, we adjust the probabilities of each state (+1, -1 and 0) so that they reflect the probability of obtaining the corresponding state in the Ruby ASI. In the Ruby ASI, considering hexagonal plaquettes on the triangular lattice, the probability of having a non-fully-formed toroidal moment (state 0) is 62/64, of having a positive fully-formed toroidal moment (state +1) is 1/64 and of having a positive fully-formed toroidal moment (state -1) is 1/64. Similarly, for triangular plaquettes on the hexagonal lattice, the probability of having a non-fully-formed toroidal moment (state 0) is 6/8, of having a positive fully-formed toroidal moment (state +1) is 1/8 and of having a positive fully-formed toroidal moment (state -1) is 1/8.

These adjusted probabilities are used to generate the initial state and any proposed state if the currents state is 0. If the current state is ±1, the probabilities of the proposed state are 1/3 for each of 0, +1 and -1. Having the same probability of 1/3 for all proposed states after a current state of ±1, is like performing loop moves in the Ruby ASI simulations, where flipping all the spins of a hexagon (or triangle) is permitted only when the associated toroidal moment is fully-formed.

**Micromagnetic simulations**

Micromagnetic simulations are conducted in Mumax3[42] using bulk material parameters for permalloy, namely a saturation magnetization of $M_{Sat} = 860$ kAm$^{-1}$, an exchange stiffness constant of $A_{ex}$=13 pJm$^{-1}$, and zero magnetocrystalline anisotropy. The damping was set to $\alpha$=0.5 to speed up convergence. The simulation cell size is set to $2.5 \times 2.5 \times 10$ nm³ in the $x$-, $y$- and $z$-directions. The $z$-direction corresponds to the thickness of the nanomagnets, and and the cell sizes in $x$ and $y$ are below the exchange length. The interaction strength between two nanomagnets is calculated as $\frac{E_1 - E_2}{2}$ where $E_1 = 2E_{nanomagnet} + E_{interaction}$ and $E_2 = 2E_{nanomagnet} - E_{interaction}$ are the total energy of the system in the two possible energy states. The vertex energies are obtained from the total energy of a vertex after relaxation to different magnetic states corresponding to the different vertex types.

**Magnetic structure factor**

The magnetic structure factor is averaged across 100 different spin configurations and is computed using the same procedure as in Ref. [43] on $1024 \times 1024$ discrete $q_x \times q_y$ points. For the Ruby ASI, the reciprocal lattice units are associated with the distances between each unit cell. This distance can be expressed as a function of the lattice parameters $(a,b)$ as $L_v = \left[\frac{a+b\sqrt{3}}{2}, \frac{\sqrt{3}a+3b}{2}\right]$. For $I_{Tr} \approx I_{Hex}$, the Bragg peaks have been artificially enlarged from a diameter of 2 points to a diameter of 10 points so that they are well-visible in the figure. The unmodified magnetic structure factor is displayed in Supplementary Figure 9.

**Effective temperature**

The effective temperature is determined by identifying the Monte Carlo simulation temperature where the vertex and flux-closed plaquette populations observed in the experimental data best match those in the simulation. Specifically, for each temperature, an error function is computed as the weighted sum of the differences between the experimental and Monte Carlo populations. In this sum, each vertex population has a weight of 1/6, while each flux-closed plaquette population has a weight of 1/2. The effective temperature is then defined as the temperature at which the error is minimised.

**Order parameters**

The ferrotoroidic order parameter $\Phi = \frac{1}{2}(|t_{H+} - t_{H-}| + |t_{T+} - t_{T-}|)$ and the Blume-Capel order parameter $\Psi = (n_{+1} + n_{-1}) + |n_{+1} - n_{-1}|$ are related and can both give a measure of the ferrotoroidic ordering in the Ruby ASI.

The ferrotoroidic order parameter $\Phi$ gives a measure of the ferrotoroidic order by subtracting the populations of positive and negative toroidal moments associated with the same shape. The absolute value of the subtraction will be highest when there are either all positive or all negative fully-formed toroidal moments. This corresponds to the ferrotoroidic ground state.

For the Blume-Capel order parameter $\Psi$ we consider only one type of toroidal moment in the Ruby ASI at a time. For example, for $a/b > a_{min}/b_{min}$ the state of the toroidal moments associated with hexagonal plaquettes is described by the Blume-Capel model variable $t$. In this case, the populations of the Blume-Capel variable $t$ in the $t = +1$ and $t = -1$ states, expressed as $n_{+1}$ and $n_{-1}$, correspond to the fractional population of hexagonal toroidal moments $t_{H+}$ and $t_{H-}$. Therefore, the order parameter $\Psi$ can be rewritten as $\Psi_{a/b \,>\, amin/bmin} = (t_{H+} + t_{H-}) + |t_{H+} - t_{H-}|$. The same approach can be applied to the toroidal moments associated with the triangular plaquettes when $a/b < a_{min}/b_{min}$.

The Blume-Capel order parameter $\Psi$ has two terms: the first term $(n_{+1} + n_{-1})$ and the second term $|n_{+1} - n_{-1}|$. The first term, $(n_{+1} + n_{-1})$, is the sum of the populations of positive and negative fully-formed toroidal moments, and is therefore the fraction of fully-formed toroidal moments present in the system. The second term, $|n_{+1} - n_{-1}|$, which is the absolute value of the difference between the populations of positive and negative toroidal moments, has the same form as the ferrotoroidic order parameter $\Phi$ and gives a measure of the ferrotoroidic alignment of toroidal moments.

In the paramagnetic phase, where there are only a few fully-formed toroidal moments, where the sign of the toroidal moments is random, both order parameters $\Phi$ and $\Psi$ are close to 0. In the paratoroidic regime, where toroidal moments of one type, either hexagonal or triangular, are fully-formed but of random sign, the ferrotoroidic order parameter $\Phi$ is close to 0, so it is not possible to distinguish between the paramagnetic and paratoroidic states. In contrast, the Blume-Capel order parameter is 1 in the paratoroidic regime because the first term has value of 1 and the second term has a value of 0. In the ferrotoroidic ground state, all toroidal moments are fully-formed and of the same sign. Therefore, the ferrotoroidic order parameter $\Phi = 1$ while the Blume-Capel order parameter $\Psi = 2$ because the both the first and the second terms have a value of 1.

Using the ferrotoroidic order parameter $\Phi$ for the Binder Cumulant analysis yields identical results to those when using the intensity of the Bragg peaks appearing at $Q_{xy} = \{\,[\,4(a + b\sqrt{3}),\; 0\,];\, [\,0, 2(\sqrt{3}a + 3b)\,]\,\}$. This confirms that the ferrotoroidic order parameter $\Phi$ is a suitable order parameter for the Ruby ASI.

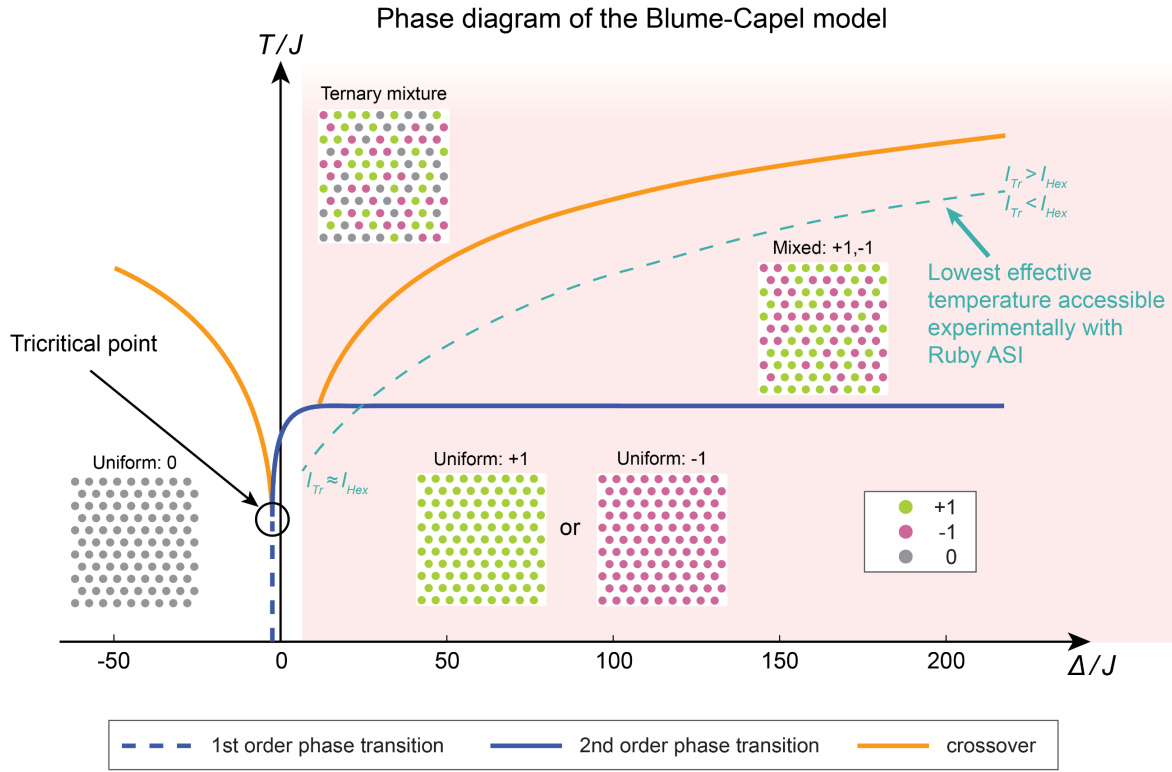

**Figure 1.** Phase diagram of the Blume–Capel model with a ferromagnetic nearest-neighbor interaction $J > 0$ and variable anisotropy $\Delta$ on a triangular lattice. The different phases are separated by either crossovers (orange lines) or phase transitions (blue lines). The phase transitions can be either first order or second order, indicated by the dashed or continuous blue lines, respectively. In the labelled insets, typical configurations in the various phases are shown. The +1, -1 and 0 spin variables are represented by green, pink and grey dots, respectively. At high temperature, there is a ternary mixture of the 0, +1 and -1 states. At intermediate temperatures and for $\Delta/J > 1$, the spins randomly assume values of $\pm 1$. At low temperatures, to the left of the tricritical point, all spins are in the 0 state, while on the right of the tricritical point there is a ferromagnetic order with all spins assuming a +1 or a -1 state. As our experimental results reveal, the region shaded in light red can be accessed using the Ruby ASI. In this region, the turquoise dashed line indicates approximately the lowest effective temperature reached in our experiments on the Ruby ASI.

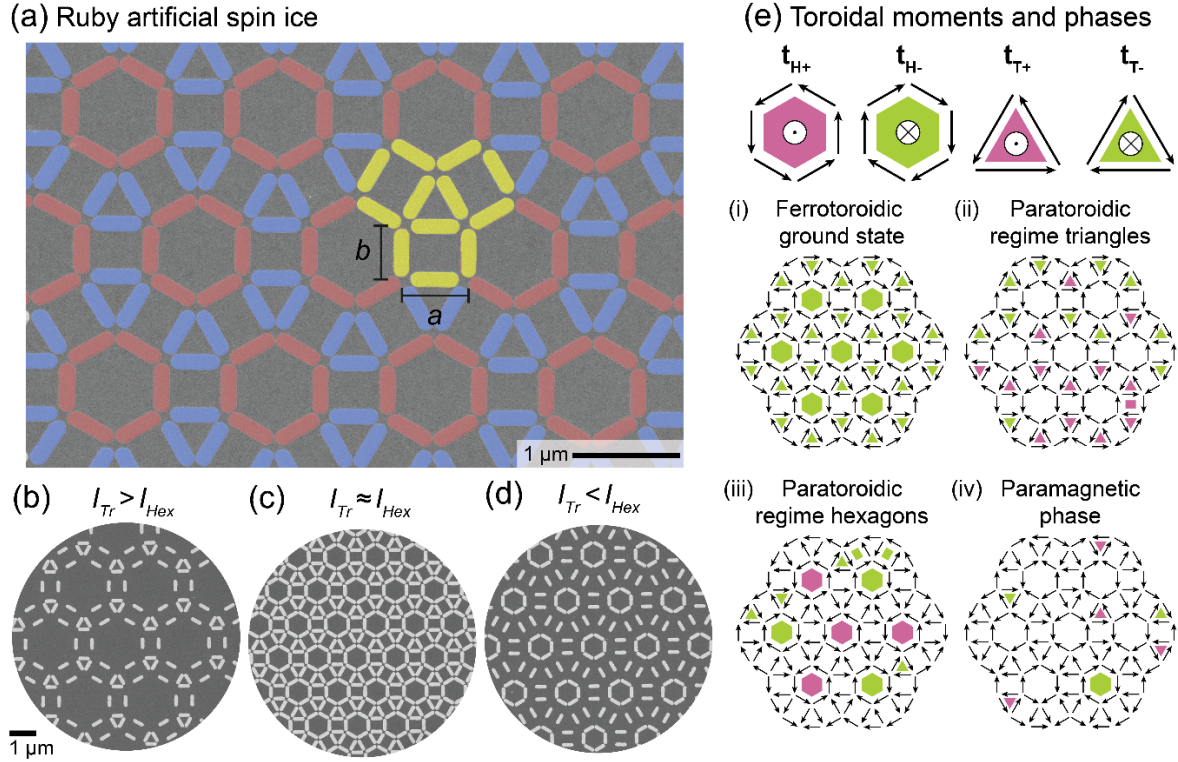

**Figure 2.** Deforming the Ruby ASI geometry to control the interactions. (a) Colored scanning electron micrograph of the Ruby ASI with the highest density of nanomagnets. Nanomagnets lining hexagons (triangles) are colored in red (blue). The unit cell, which contains twelve nanomagnets, is highlighted in yellow. The lattice parameters a and b control the side lengths of the triangles and hexagons, respectively. (b)-(d) Scanning electron micrographs of the Ruby ASI for the three cases: $I_{Tr} > I_{Hex}$, $I_{Tr} \approx I_{Hex}$ and $I_{Tr} < I_{Hex}$. (e) Schematics of positive and negative toroidal moments in flux-closed loops of macrospins in triangular and hexagonal plaquettes (top row), and example states representing the phases and regimes in the Ruby ASI (lower four panels). The macrospins are indicated with black arrows.

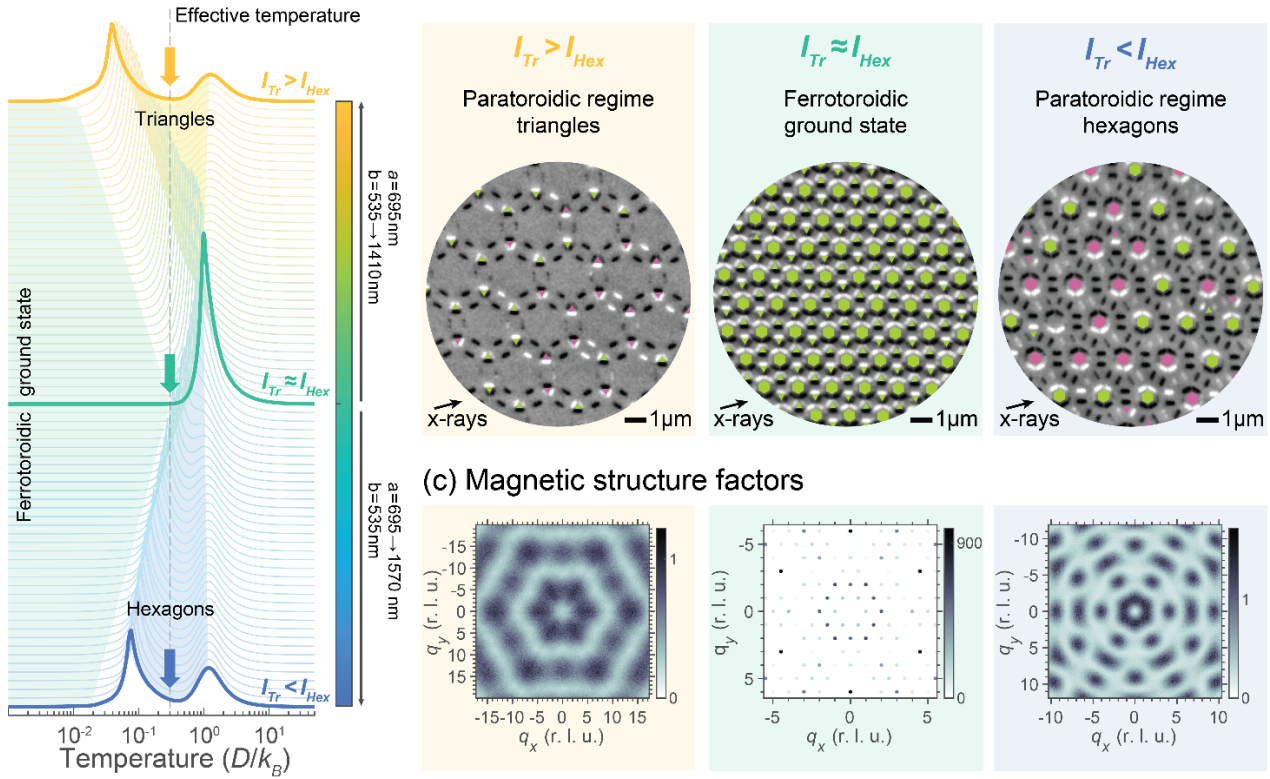

**Figure 3.** Going from one-step to two-step ordering in the Ruby ASI by changing $a/b$. (a) Waterfall plot of the heat capacity as a function of temperature for 72 different pairs of lattice parameters, a and b. The colours reflect the lattice parameter ratio a/b, which is obtained by: keeping a constant at 695 nm and varying b from 1410 nm to 535 nm when going from the top to the middle, and varying a from 695 nm to 1570 nm while keeping b constant at 535 nm when going from the middle to bottom. Towards the top and bottom of the waterfall plot, there are two peaks in heat capacity while, in the middle, only one peak is featured. The vertical grey dashed line indicates the approximate effective temperature reached by thermal annealing. (b) PEEM images with fully-formed toroidal moment indicated by green and pink coloured plaquettes, for the three cases $I_{Tr} > I_{Hex}$, $I_{Tr} \approx I_{Hex}$ and $I_{Tr} < I_{Hex}$, which have magnetic configurations in the paratoroidic regime triangles, the ferrotoroidic ground state and the paratoroidic regime hexagons, respectively. (c) Magnetic structure factor of spin configurations from the Monte Carlo simulation, at the effective temperature reached by thermal annealing, for the three cases of $I_{Tr} > I_{Hex}$, $I_{Tr} \approx I_{Hex}$ and $I_{Tr} < I_{Hex}$.

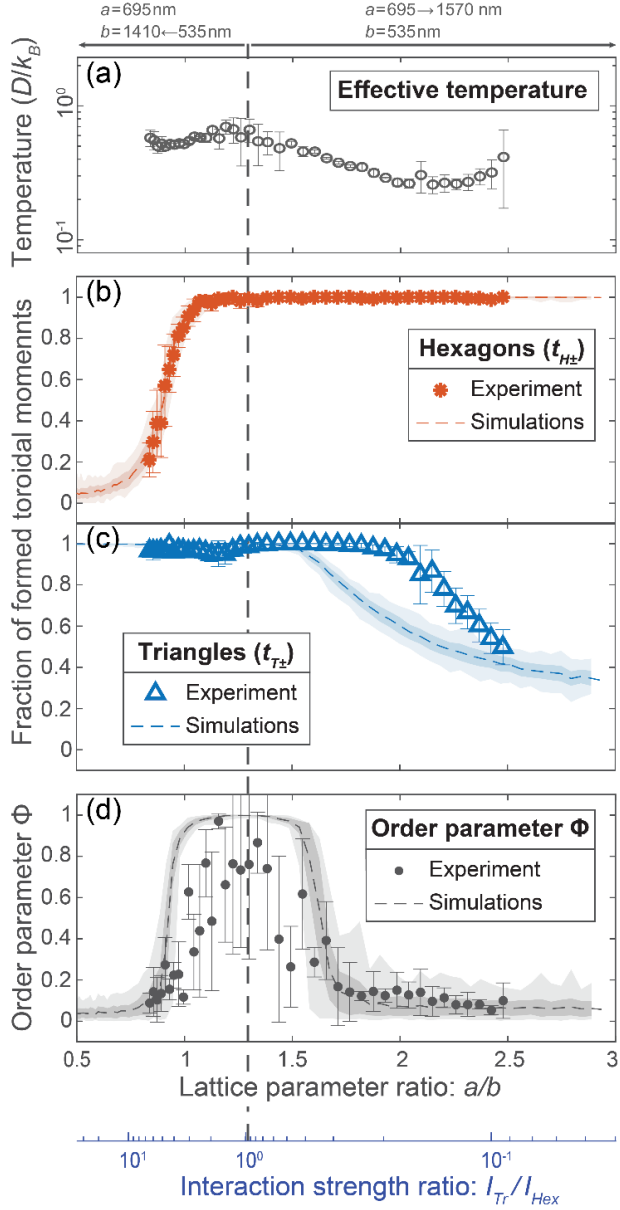

**Figure 4.** Signatures of the two-step ordering process and toroidic phases in as-grown configurations determined with MFM. (a) Average effective temperature of as-grown states as a function of the lattice parameter ratio. (b)-(c) Average fraction of formed toroidal moments associated with (b) hexagons $t_{H\pm} = t_{H+} + t_{H-}$ and (c) triangles $t_{T\pm} = t_{T+} + t_{T-}$ in as-grown states as a function of the lattice parameter ratio. (d) Average ferrotoroidic order parameter $\Phi = \frac{1}{2}(|t_{H+} - t_{H-}| + |t_{T+} - t_{T-}|)$ as a function of the lattice parameter ratio. For panels (b)-(d), the dashed line indicates average values obtained from 100 individual Monte Carlo simulations at a temperature $k_B T/D \approx 0.469$. The inner coloured area indicates the standard deviation across 100 individual simulations while the outer lightly coloured area indicated the maximum and minimum values. For all the graphs, each data point is the average of the represented quantity as extracted from four MFM measurements for each lattice parameter of as-grown samples. The error bars indicate the standard deviation of the represented quantity.

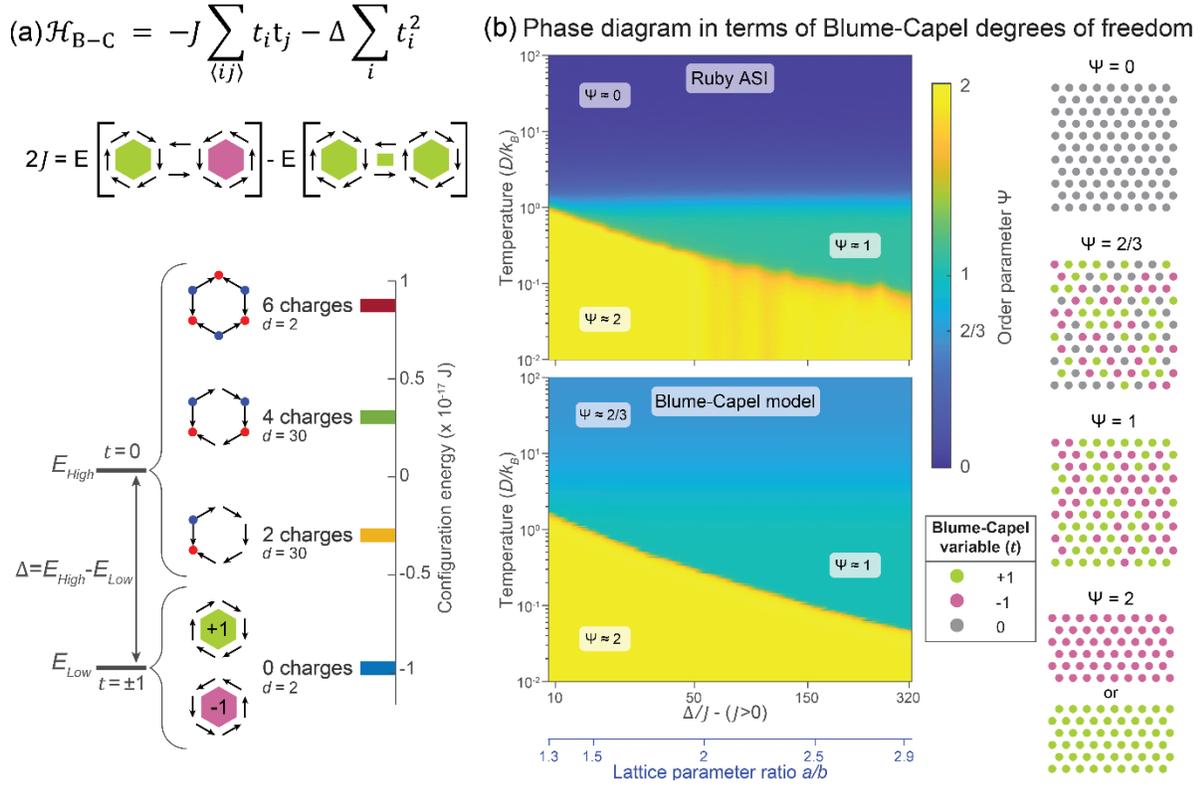

**Figure 5.** Assigning Blume-Capel degrees of freedom to the Ruby ASI. (a) The Hamiltonian of the Blume-Capel model is composed of two terms: a nearest-neighbour interaction of strength J, and an anisotropy term Δ, which is the energy difference between the 0 and the ±1 states. In our effective model for the Ruby ASI, the nearest-neighbour interaction J is given by the energy difference between a ferrotoroidic ground state and a configuration where one toroidal moment is reversed. The quantity Δ is defined as the energy gap between the average energy of excited states ($E_{High}$) and the ground state ($E_{Low}$) of a hexagonal plaquette. The blue and red dots indicate the position of positive and negative charges, respectively. For each possible state, we indicate the number of charges and the degeneracy $d$. (b) Color plots of the order parameter $\Psi = (n_{+1} + n_{-1}) + |n_{+1} - n_{-1}|$ obtained from Monte Carlo simulations of the Ruby ASI (upper panel) and Blume-Capel model (lower panel) as a function of temperature and Δ/J, which is dictated by the lattice parameter ratio a/b. The quantities $n_{+1}$ and $n_{-1}$ are the fractional populations of positive and negative toroidal moments, respectively. At high temperature $\Psi \approx 2/64$ for the Ruby ASI with few fully-formed toroidal moments and $\Psi \approx 2/3$ for the Blume-Capel model which reflects the proportion of +1 and -1 states, with all states (0, +1 and -1) having equal probabilities of 1/3. Below the crossover, controlled by Δ, the system is in the paratoroidic phase, with $\Psi \approx 1$, where all toroidal moments are formed but have random values. In the low temperature ferrotoroidic phase, $\Psi = 2$ because all the toroidal moments are aligned, with all either +1 or all -1.

**Present Addresses**

†Present address: Department of Physics, Princeton University, Princeton, NJ 08540 USA.

**Author Contributions**

L.B, G.M.M and L.J.H conceived the project. L.B. designed and fabricated the samples, and took the MFM data. L.B., G.M.M., T.W and A.K. took the PEEM measurements. L.B. and G.M.M. wrote the Monte Carlo code. L.B analysed the data and, with the assistance of G.M.M., F.M and P.M.D., interpreted the results. L.B., G.M.M. and L.J.H wrote the manuscript with input from the other authors.

ACKNOWLEDGMENT

This work was funded by the Swiss National Science Foundation (project no. 200020_200332). Part of this work was performed on the Merlin6 High Performance Computing Cluster and the Surfaces / Interfaces: Microscopy (SIM) beamline of the Swiss Light Source at the Paul Scherrer Institute, Villigen.

**SUPPLEMENTARY FIGURES**

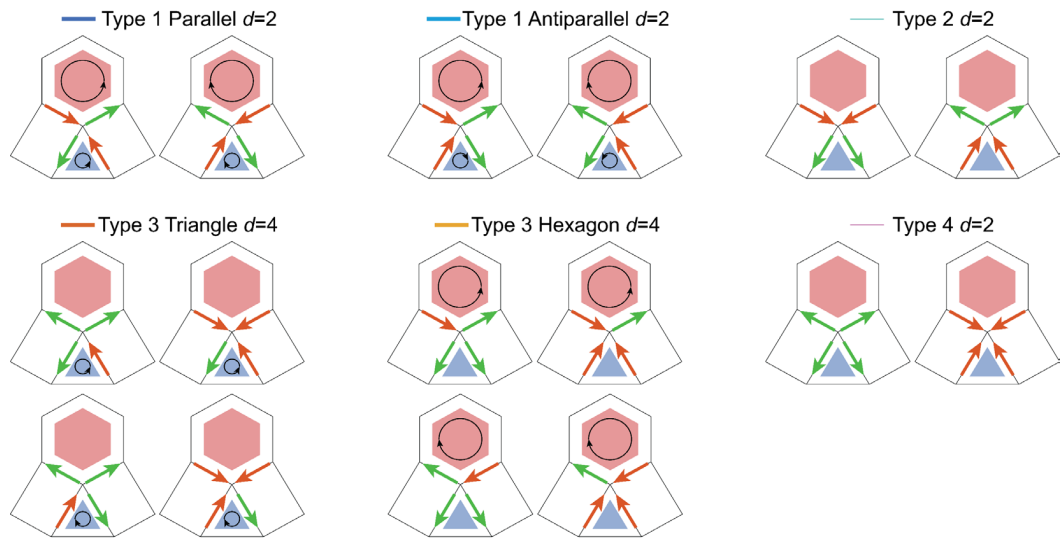

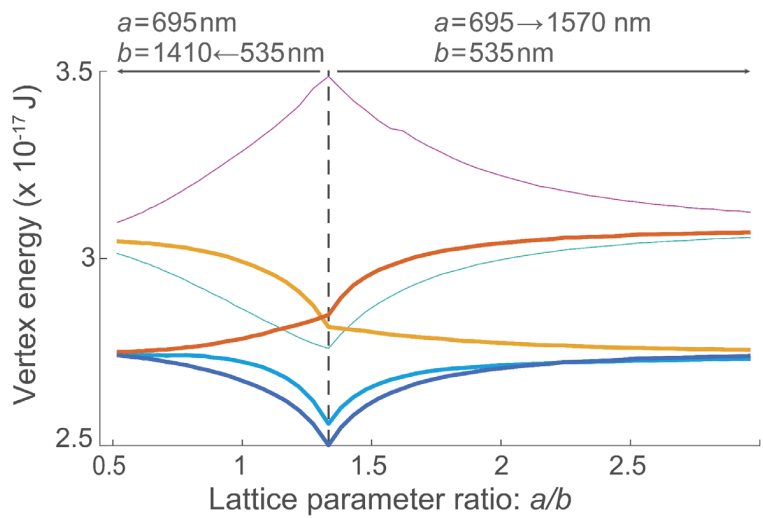

**Figure S1.** Vertex types and energies in the Ruby ASI. All vertex configurations present in the Ruby artificial spin ice are grouped into six vertex types. The degeneracy d of each vertex type is indicated after the name of the vertex type. The vertex configurations are displayed over a schematic outline of the lattice (black lines). Within the schematic outline of the lattice, the hexagonal plaquettes are indicated by a red hexagon and the triangular plaquettes by a blue triangle. If the nanomagnets along the edge of a triangle or a hexagon are aligned head-to-tail, the sense of circulation is indicated by a black circular arrow. Below: vertex energies as a function of the lattice parameter ratio a/b. The graph has two parts as a function of a/b: starting at the left to the point marked by the dashed line, the lattice parameter a is kept constant at 695 nm while b is varied from 1410 nm to 535 nm; then, from the dashed line to the right of the figure, the lattice parameter a is varied from 695 nm to 1570 nm and b is kept constant at 535 nm.

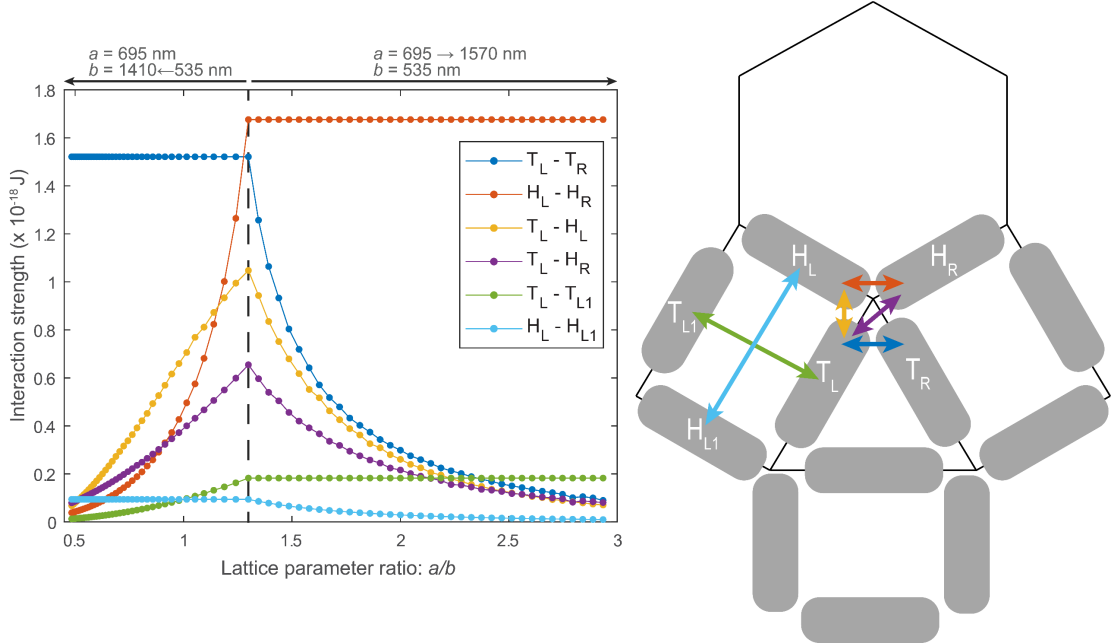

**Figure S2.** Pairwise interactions from micromagnetic simulations. Pairwise interaction strengths as a function of the lattice parameter ratio a/b for all nanomagnet pairs within a unit cell. The graph has two parts: going from the left to the vertical dashed line, the lattice parameter a is kept constant at 695 nm and b is varied from 1410 nm to 535 nm; going from the vertical dashed line to the right, the lattice parameter a is varied from 695 nm to 1570 nm and b is kept constant at 535 nm. The definitions of the nanomagnet pairs are shown on the right over an annotated unit cell of the Ruby lattice ASI. We label nanomagnets T or H if they belong to a triangular or a hexagonal plaquette, respectively. The subscripts L and R denote whether a nanomagnet is on the left or the right with respect to the centre of unit cell, while the subscript 1, if present, indicates that the nanomagnet belongs to a different triangular or hexagonal plaquette than the one present in the unit cell schematic. In the main text, the interaction energy between nanomagnets $T_L$ and $T_R$ is called $I_{Tr}$, while the interaction energy between nanomagnets $H_L$ and $H_R$ is called $I_{Hex}$.

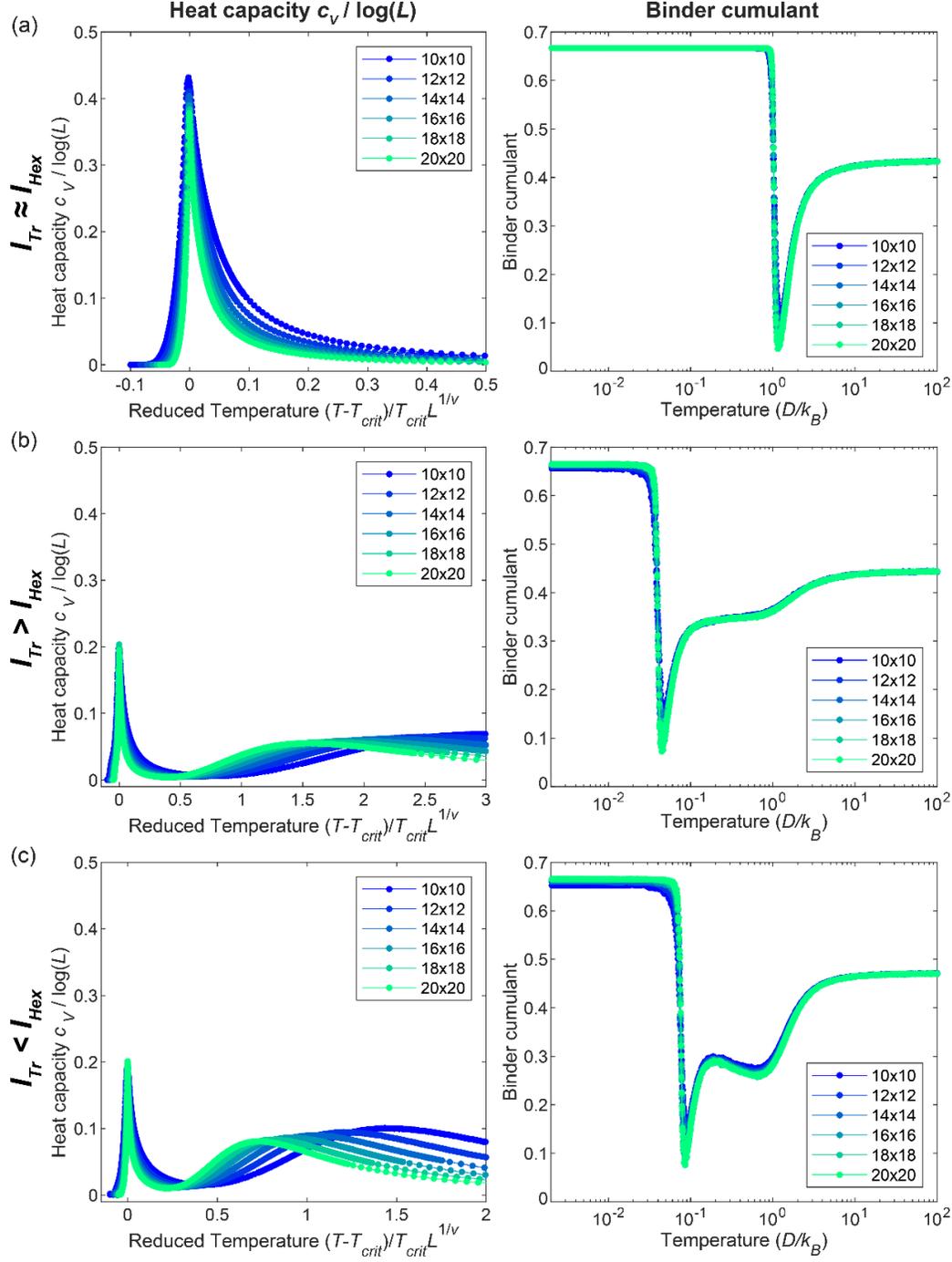

**Figure S3.** Monte Carlo simulations and scaling behaviour of the Ruby ASI. On the left the scaled heat capacity, $c_V / \log(L)$, is plotted against the reduced temperature which has been obtained from the Binder 4th order cumulant analysis. There curves appear to collapse when assuming the relevant critical exponent for a 2D Ising transition, namely $\nu = 1$. On the right, the Binder cumulant is plotted as a function of temperature. The plots are for a range of systems sizes $L$ from $10 \times 10$ to $20 \times 20$ unit cells. The finite size scaling analysis has been performed on the three extreme cases of $I_{Tr} \approx I_{Hex}$ ($a = 695$ nm, $b = 535$ nm) in panel (a), $I_{Tr} > I_{Hex}$ ($a = 695$ nm, $b = 1410$ nm) in panel (b) and $I_{Tr} < I_{Hex}$ ($a = 1570$ nm, $b = 535$ nm) in panel (c).

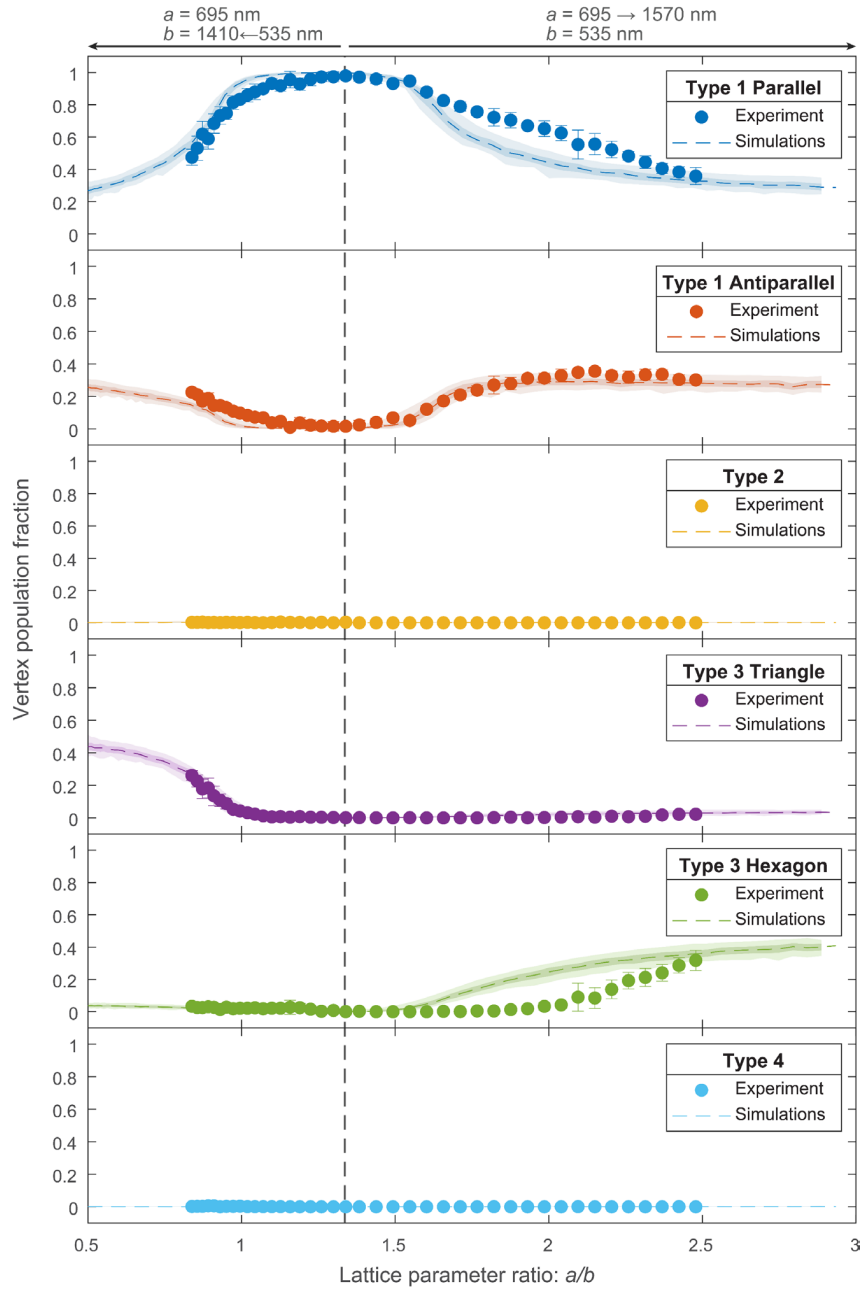

**Figure S4.** Vertex populations in as-grown configurations as a function of the lattice parameter ratio. For all the graphs, each data point is the average of the represented quantity as extracted from four MFM measurements for each lattice parameter of as-grown samples and the error bars indicate the standard deviation. The dashed line indicates average values obtained from 100 individual Monte Carlo simulations at a temperature $k_B T/D \approx 0.469$. The inner coloured area indicates the standard deviation while the outer lightly-coloured area indicates the maximum and minimum values. The graph has two parts: going from the left to the vertical dashed line, the lattice parameter $a$ is kept constant at 695 nm and $b$ is varied from 1410 nm to 535 nm; from the vertical dashed line to the right, the lattice parameter $a$ is varied from 695 nm to 1570 nm and $b$ is kept constant at 535 nm.

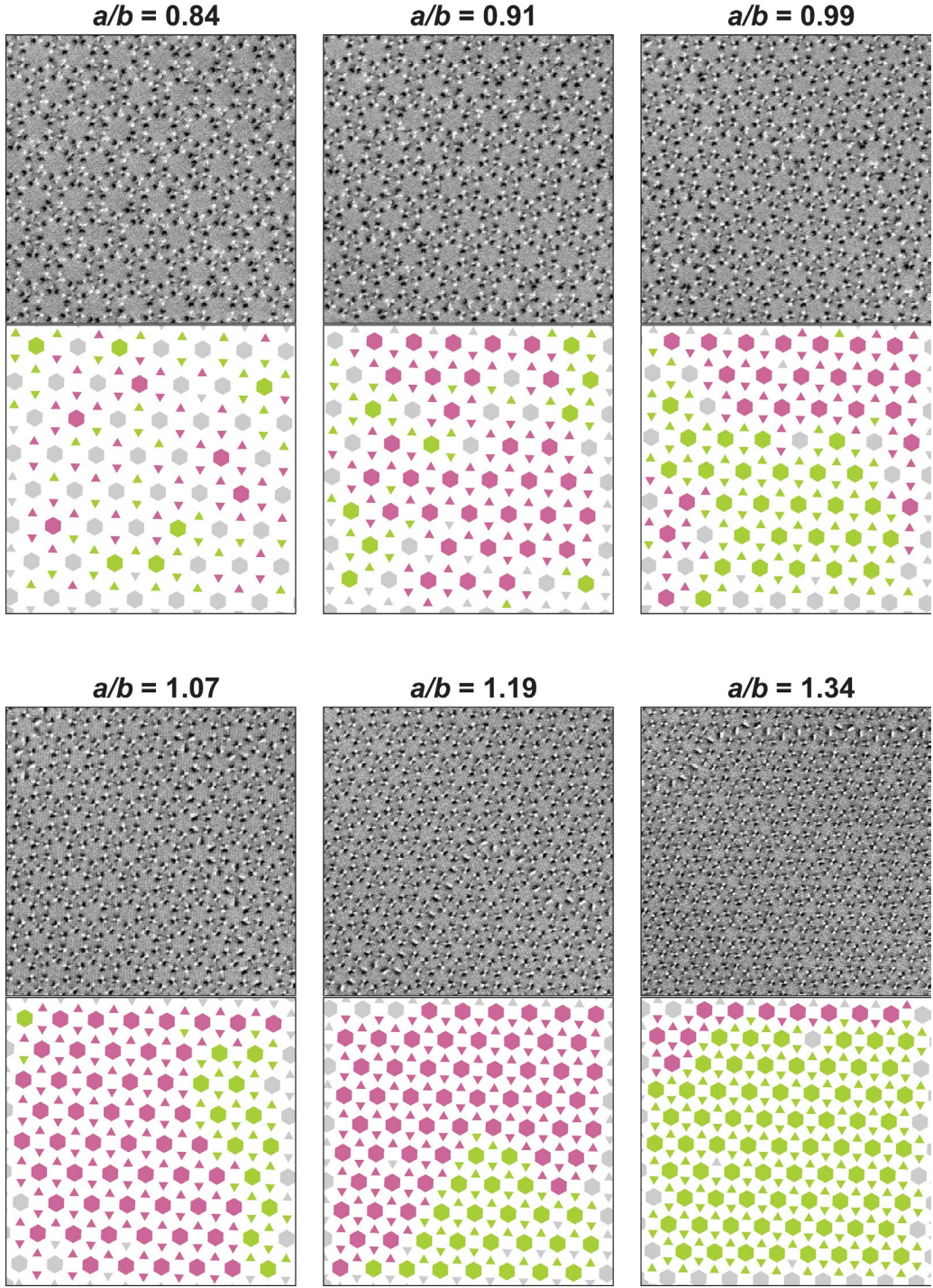

**Figure S5.** Ruby ASI as grown configurations for increasing *a/b* – part I. For selected *a/b*, from 0.84 up to 1.34, MFM images and toroidal moment maps of as grown configurations are given. Pink triangular and hexagonal plaquettes represent positive fully-formed toroidal moments. Green triangular and hexagonal plaquettes represent negative fully-formed toroidal moments. Grey triangular and hexagonal plaquettes indicate plaquettes that do not have a fully-formed toroidal moment or, equivalently, where not all of the macrospins point head-to-tail.

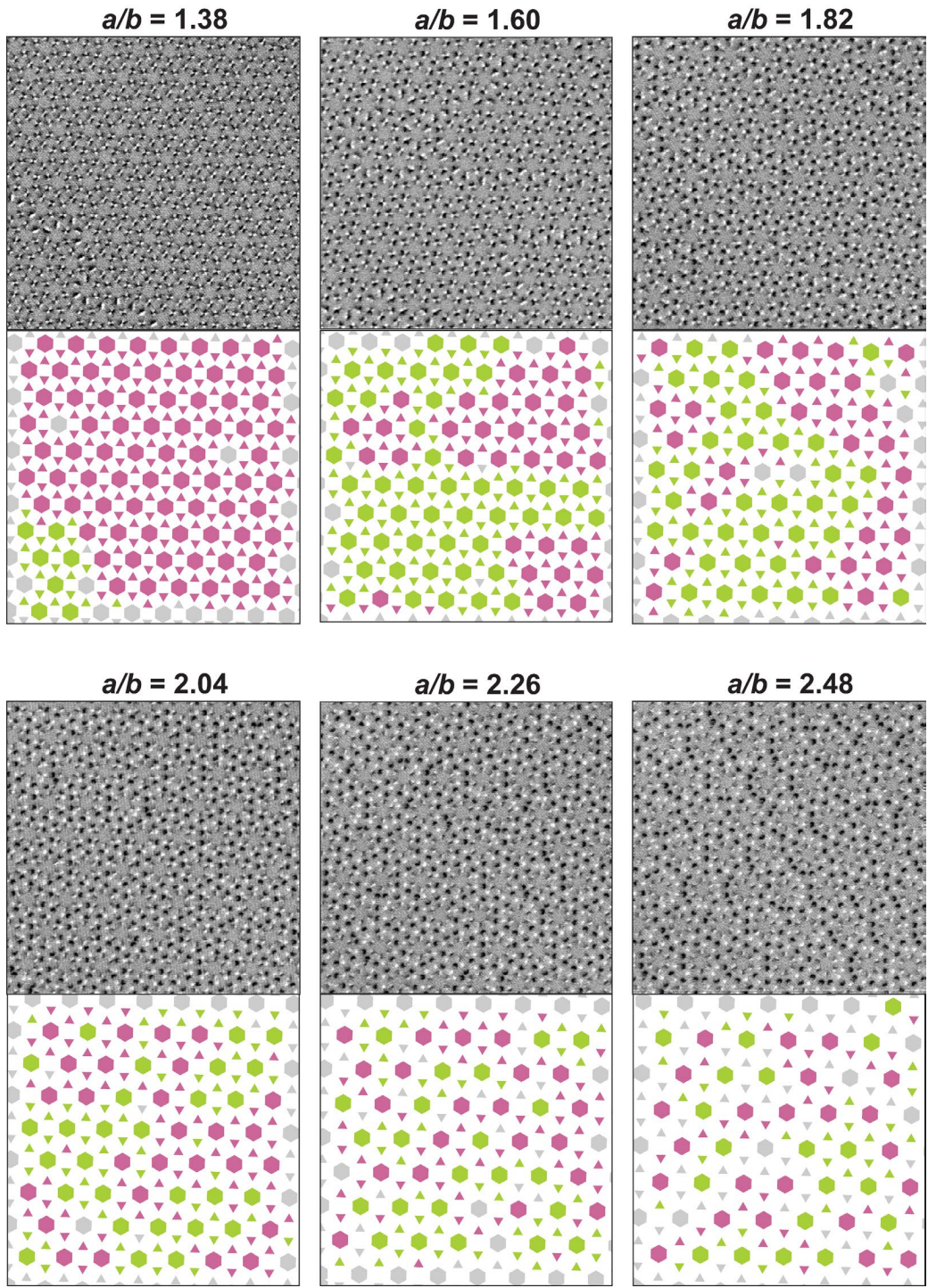

**Figure S6.** Ruby ASI as grown configurations for increasing *a/b* – part II. For selected *a/b*, from 1.38 up to 2.48, MFM images and toroidal moment maps of as-grown configurations are given. Pink triangular and hexagonal plaquettes represent positive fully-formed toroidal moments. Green triangular and hexagonal plaquettes represent negative fully-formed toroidal moments. Grey triangular and hexagonal plaquettes indicate plaquettes that do not have a full-formed toroidal moment or, equivalently, where not all of the macrospins point head-to-tail.

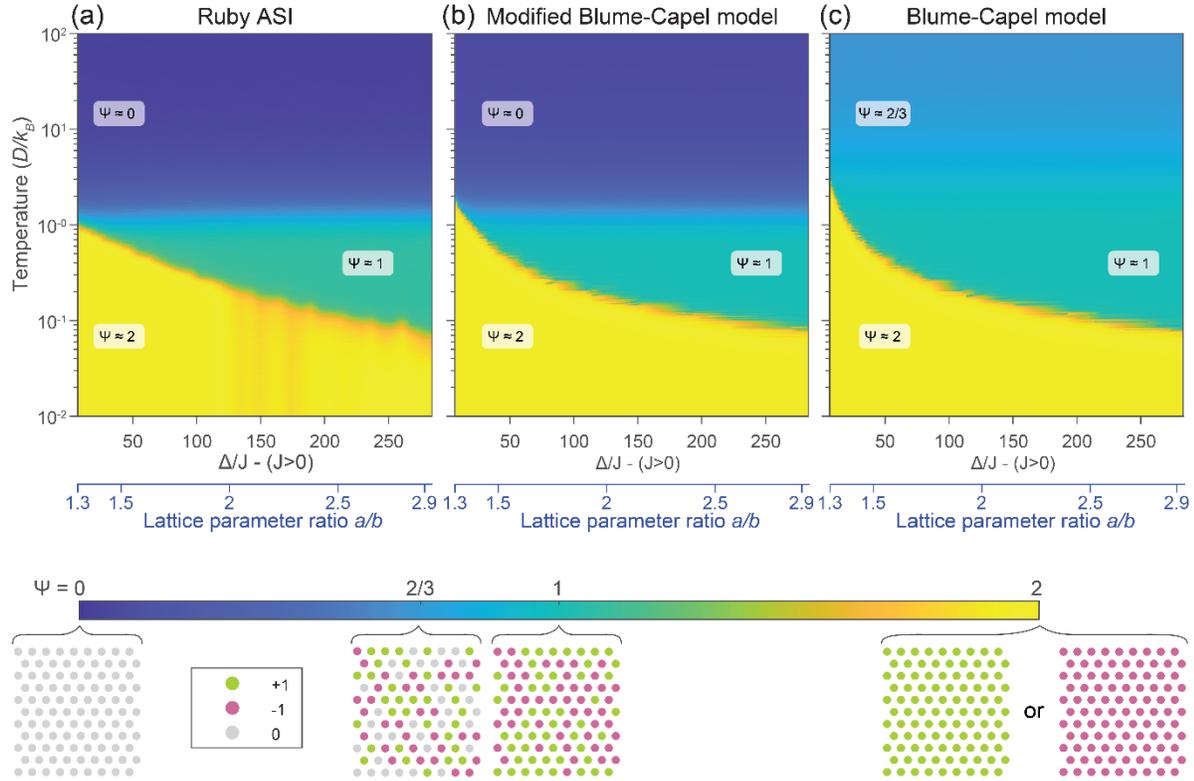

**Figure S7.** Blume-Capel model for hexagonal toroidal moments on a triangular lattice. Color plots of the order parameter $\Psi = (n_{+1} + n_{-1}) + |n_{+1} - n_{-1}|$ as function of temperature for: (a), the Ruby ASI, , where the full spin ensemble is considered and coupled through point-dipolar interactions; (b), the modified Blume-Capel model; and c, the Blume-Capel model as a function of temperature and $\Delta/J$, which is dictated by the lattice parameter ratio a/b. The quantities $n_{+1}$ and $n_{-1}$ are the populations of positive and negative toroidal moments, respectively. In (a) and (b), $\Psi \approx 2/64$ at high temperature (dark blue region) since there are few fully-formed toroidal moments, while in (c), $\Psi \approx 2/3$ at high temperature (light blue region) since the toroidal moments are generated to be 0, +1 or -1 with equal probabilities. In all panels, below the crossover controlled by $\Delta$, the system is in the paratoroidic phase, with $\Psi \approx 1$ (green region), where all toroidal moments have formed but have random values. The low temperature ferrotoroidic phase, with $\Psi = 2$ (yellow region), has all toroidal moments aligned with values of +1 or -1.

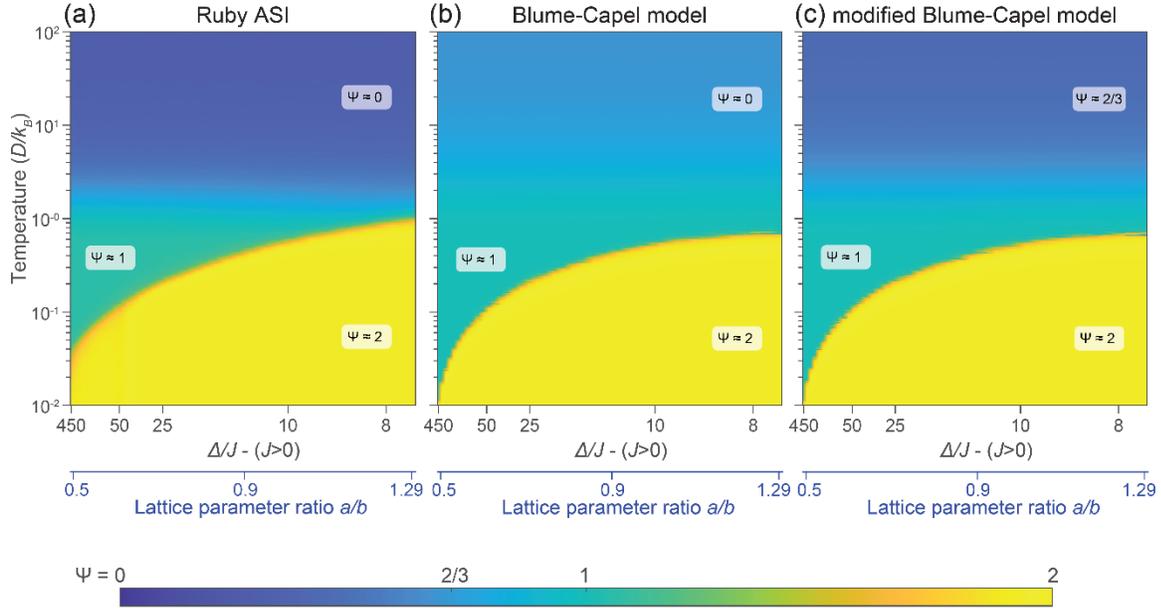

**Figure S8.** Blume-Capel model for triangular toroidal moments on a hexagonal lattice. Colour plots of the order parameter $\Psi = (n_{+1} + n_{-1}) + |n_{+1} - n_{-1}|$ as a function of temperature for: (a), the Ruby ASI, where the full spin ensemble is considered and coupled through point-dipolar interactions; (b), the Blume-Capel model; and **c**, the modified Blume-Capel model. The Monte Carlo simulations of the Blume-Capel models are performed in terms of $\Delta/J$, which is dictated by the lattice parameter ratio a/b. The quantities $n_{+1}$ and $n_{-1}$ are the populations of fully-formed toroidal moments with value +1 and -1 respectively. In (a) and (c), $\Psi \approx 2/8$ at high temperatures, giving a dark blue region, since there are only a few formed toroidal moments, while in (b) $\Psi \approx 2/3$ at high temperature, giving a light blue region, since the toroidal moments are 0, +1 or -1 with equal probabilities. In all panels, below the crossover controlled by $\Delta$, the system is in the paratoroidic phase, with $\Psi \approx 1$ (green region), since all toroidal moments are formed but have random values. The low temperature ferrotoroidic phase with $\Psi = 2$ (yellow region), has all toroidal moments aligned with values of +1 or -1.

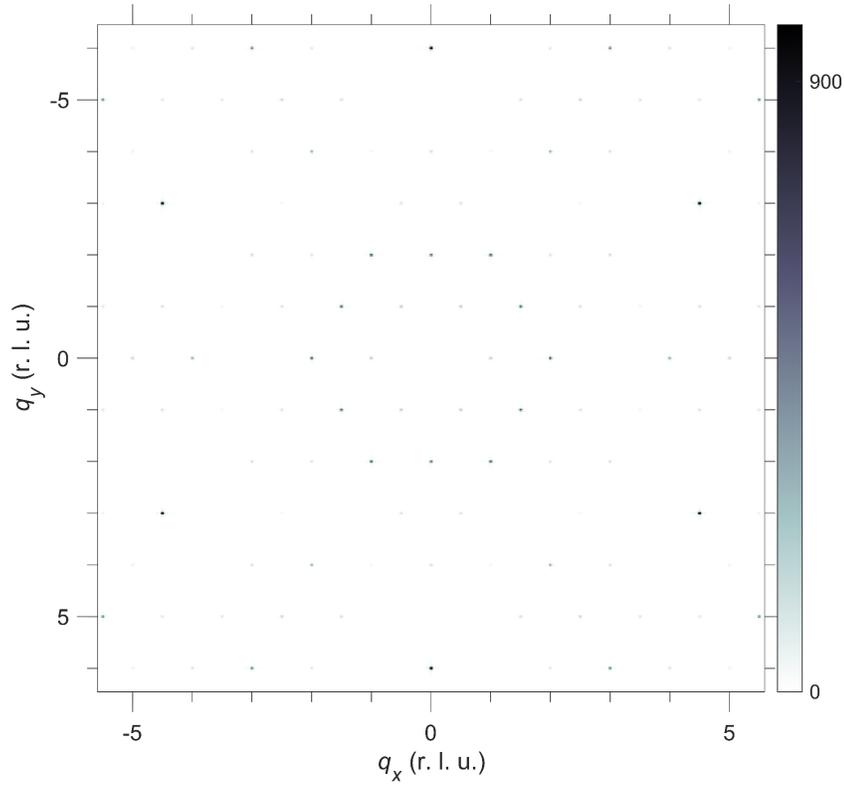

**Figure S9.** Magnetic structure factor for $I_{Tr} \approx I_{Hex}$. Magnetic structure factor averaged over 100 independent spin configurations from the Monte Carlo simulation, at the effective temperature reached by thermal annealing, for the case where $I_{Tr} \approx I_{Hex}$. In this figure, the original Bragg peaks are shown at their true scale in terms of number of pixels. This contrasts with the middle panel of Figure. 3c in the main text, where the peaks were artificially enlarged due to space constraints to ensure they were visible.